\documentclass[aps,prl,twocolumn,superscriptaddress,footnotebib]{revtex4-1}
\usepackage{graphicx}
\usepackage{amssymb,amsmath}
\usepackage{bm}
\usepackage{dcolumn}
\usepackage{subfigure}
\usepackage{float}
\usepackage[OT1]{fontenc} 
\usepackage{url}
\usepackage{slashed}
\usepackage{color}
\usepackage{verbatim}
\usepackage{txfonts}
\usepackage[driverfallback=dvipdfm]{hyperref}
\hypersetup{pdfpagemode=FullScreen,colorlinks=true,breaklinks,urlcolor=blue,linkcolor=blue,citecolor=blue}
\usepackage{natbib}
\usepackage{verbatim}

\usepackage{subfigure}
\usepackage{comment}
\usepackage{hyperref}
\usepackage{amssymb}
\usepackage{bm}
\usepackage{graphicx}
\usepackage{amsmath,amssymb}
\usepackage{tikz,fp}
\usepackage{tikz-cd}
\usetikzlibrary{arrows}
\usetikzlibrary{intersections}
\usetikzlibrary{shapes.geometric}
\usetikzlibrary{decorations.pathmorphing, patterns,shapes,fixedpointarithmetic}
\usetikzlibrary{decorations.markings}

\pgfdeclarepatternformonly{south west lines}{\pgfqpoint{-0pt}{-0pt}}{\pgfqpoint{3pt}{3pt}}{\pgfqpoint{3pt}{3pt}}{
	\pgfsetlinewidth{0.4pt}
	\pgfpathmoveto{\pgfqpoint{0pt}{0pt}}
	\pgfpathlineto{\pgfqpoint{3pt}{3pt}}
	\pgfpathmoveto{\pgfqpoint{2.8pt}{-.2pt}}
	\pgfpathlineto{\pgfqpoint{3.2pt}{.2pt}}
	\pgfpathmoveto{\pgfqpoint{-.2pt}{2.8pt}}
	\pgfpathlineto{\pgfqpoint{.2pt}{3.2pt}}
	\pgfusepath{stroke}}

\tikzset{
	mid arrow/.style={postaction={decorate,decoration={
				markings,
				mark=at position .575 with {\arrow{stealth}}
	}}},
	near arrow/.style={postaction={decorate,decoration={
				markings,
				mark=at position .275 with {\arrow{stealth}}
	}}},
	far arrow/.style={postaction={decorate,decoration={
				markings,
				mark=at position .800 with {\arrow{stealth}}
	}}},
	snake arrow/.style={fixed point arithmetic, decorate, decoration={snake,amplitude=2pt, segment length=11pt},postaction={decoration={markings,mark=at position 0.625 with {\arrow{stealth}}},decorate}},
}

\begin{document}
	
	\title{Boundary Condition Independence of Non-Hermitian Hamiltonian Dynamics }
	\author{Liang Mao}
	\thanks{They contribute equally to this work.}
	\affiliation{Institute for Advanced Study, Tsinghua University, Beijing 100084, China}
	\affiliation{Department of Physics, Tsinghua University, Beijing 100084, China}

	\author{Tianshu Deng}
	\thanks{They contribute equally to this work.}
	\affiliation{Institute for Advanced Study, Tsinghua University, Beijing 100084, China}
	
	\author{Pengfei Zhang}
	\thanks{pengfeizhang.physics@gmail.com}
	\affiliation{Walter Burke Institute for Theoretical Physics, California Institute of Technology, Pasadena, CA 91125, USA}
	\affiliation{Institute for Quantum Information and Matter, California Institute of Technology, Pasadena, CA 91125, USA}
	\date{\today}
	
	\begin{abstract}
    Non-Hermitian skin effect, namely that the eigenvalues and eigenstates of a non-Hermitian tight-binding Hamiltonian have significant differences under open or periodic boundary conditions, is a remarkable phenomenon of non-Hermitian systems. Inspired by the presence of the non-Hermitian skin effect, we study the evolution of wave-packets in non-Hermitian systems, which can be determined using the single-particle Green's function. Surprisingly, we find that in the thermodynamical limit, the Green's function does not depend on boundary conditions, despite the presence of skin effect. We proffer a general proof for this statement in arbitrary dimension with finite hopping range, with an explicit illustration in the non-Hermitian Su-Schrieffer-Heeger model. We also explore its applications in non-interacting open quantum systems described by the master equation, where we demonstrate that the evolution of the density matrix is independent of the boundary condition. 
	\end{abstract}
	
	\maketitle
  {\color{blue}\emph{Introduction.}} --   
  In recent years, non-Hermitian physics \cite{NH1,NH2} has arrested a lot of attention in both classical and quantum physics. In classical physics, for example, a non-Hermitian effective Hamiltonian can describe photonic or acoustic systems with loss and gain \cite{optics1,optics2,optics3,acoustic1}. In quantum physics, the non-Hermiticity can be introduced by making the system coupled with an external bath \cite{open1,open2,open3,open4,NSHEopen1,open5,pan2020non}. The non-Hermiticity can lead to novel physical phenomena. As an example, when tuning parameters, a non-Hermitian Hamiltonian can have Exceptional Points where two or more eigenvalues and eigenstates coalesce \cite{EP1,EP2,EP3,EP4,EP5,EP6,EP7,EP8}.
  
  Among these innovative studies, the interplay between non-Hermiticity and topology attracts much attention both theoretically \cite{NHtopoThe1,NHtopoThe2,NHtopoThe3,NHtopoThe4,NHtopoThe5,NHtopoThe6,NHtopoThe7,NHtopoThe8,NHtopoThe9,NHtopoThe10,NSHE1,NSHE2,NSHE5,NSHE7,nonssh1,nonssh3,chen2018hall} and experimentally \cite{NHtopoExp1,NHtopoExp2,NHtopoExp3,NHtopoExp4,NHtopoExp5,NHtopoExp6,NHtopoExp7,NHtopoExp8,NHtopoExp10,NHtopoExp11,NHtopoExp12,NHtopoExp13,NHtopoExp14,NHtopoExp15,NHtopoExp16}. Conventionally, the topological phenomena are closely related to the bulk topological invariants defined on the Brillouin zone, through the bulk-boundary correspondence \cite{topo1,topo2,topo3,topo4,topo5,topo6,topo7,topo8}. In non-Hermitian systems, however, the topological invariants are defined on the generalized Brillouin zone (GBZ) \cite{NSHE1,nonblochband1,nonblochband2}. This is due to the celebrated non-Hermitian skin effect (NHSE) \cite{NSHE1,NSHE3,NSHE2,NSHE4,NSHE5,NSHE6,NSHE7,NSHEopen1,NSHE9,NSHE8dim,NSHE10}, which states that the majority of eigenstates of the non-Hermitian Hamiltonian with the open boundary condition are exponentially localized at boundaries. On the contrary, when the periodic boundary condition is imposed, the eigenstates are plane waves modulated by the periodic potential, as implied by Bloch's theorem \cite{kittel1996introduction}. Subsequently, the breakdown of Bloch's theorem under the open boundary condition suggests a boundary sensitivity of eigenvalues and eigenstates, even if we take the thermodynamical limit. 
\begin{figure}[tb]
		\centering
		\includegraphics[width=0.95\linewidth]{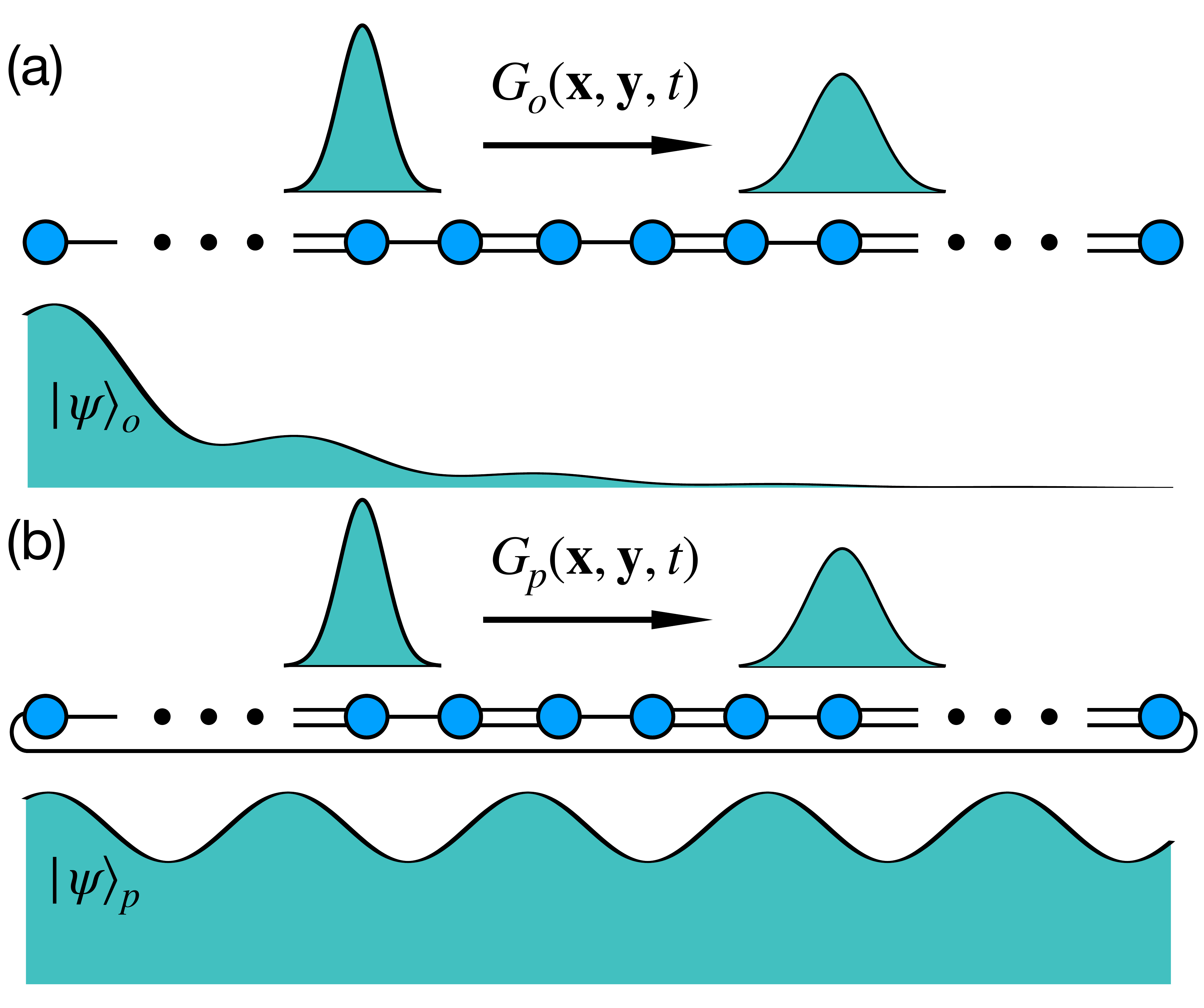}
		\caption{Under (a) the open and (b) the periodic boundary condition, the wave function $|\psi\rangle$ behaves in different manners in the thermodynamic limit. However, the single particle propagator $G_{o/p}(\mathbf{x},\mathbf{y},t)$ remains the same, thus generating the same bulk dynamics. Here the subscript indicates the boundary condition.}
		\label{fig:1}
	\end{figure}

  The sensitivity of eigenstates and eigenvalues seems to imply that the evolution of a wave packet under the open boundary condition should be different from the evolution under the periodic boundary condition, even in the thermodynamical limit. However, in this work, we give a general proof that the single-particle Green's function $G(t)=\left<f\right|e^{-iHt}\left|i\right>$, and thus the evolution of the wave packet, is independent of the boundary condition in the thermodynamical limit. 
  Our proof works for general dimensions with finite hopping range and number of bands. We then give an explicit example for the equivalence using the non-Hermitian Su-Schrieffer-Heeger (SSH) model \cite{NSHE1,NSHE3,nonssh1,nonssh2,nonblochband1,nonssh3}. In this case, the Green's function can be analytically reduced to a contour integral, and the Green's function under the open boundary condition can be shown the same as its close boundary condition counterpart by a contour deformation. Finally, we apply our results to the open quantum systems, where we prove that the evolution under quadratic Master equations is insensitive to the boundary condition.

  {\color{blue}\emph{Model.}} -- We consider general non-interacting non-Hermitian systems in the $D$-dimension, with the Hamiltonian 
\begin{equation}\label{Hamiltonian}
    \hat H =\sum_{\mathbf{x,y}}H_{\mathbf{x}\mathbf{y}}^{\mu\nu}\hat{c}^\dagger_{\mathbf x,\mu}\hat{c}_{\mathbf y,\nu}= \sum_{\mathbf{x}}\sum_{\text{max}\{r_i\}<N}\sum_{\mu,\nu=1}^q t_{\mathbf{r},\mu\nu}\hat{c}^\dagger_{\mathbf x+\mathbf{r},\mu}\hat{c}_{\mathbf x,\nu}
\end{equation}
Here $\mathbf{x}=(x_1,x_2...,x_D)$ labels different sites and $\mathbf{r}=(r_1,r_2...,r_D)$ labels the displacement. $\mu,\nu=1,2...q$ labels different sites in a unit cell and $N$ is the range of hopping. To be concrete, we consider a system with $x_i \in \{1,2,...,L\}$ with the periodic boundary conditions in $x_2$, ... $x_D$. If we take an open boundary condition in $x_1$, the single-particle wavefunction of eigenstates then takes the form of 
$$\psi_{\mathbf{x},\mu}(\mathbf{k}_\perp)=\sum_{a=1}^{2M} (\beta_a)^{x_1} e^{i\mathbf{k}_\perp\cdot \mathbf{x}_\perp} \phi_\mu^a(\mathbf{k}_\perp).$$
Here $M=qN$ and we have defined the $D-1$ dimensional momentum $\mathbf{k}_\perp=(k_2,...,k_D)$. $\beta_a$ is determined by both the eigenequations in bulk and the boundary condition $|\beta_M|=|\beta_{M+1}|$, which generally leads to $|\beta_a|\neq1$. Consequently, the wavefunction is exponentially localized around the boundary, known as the non-Hermitian skin effect. The allowed value of $\beta_M$ and $\beta_{M+1}$ form a close cycle at the thermodynamic limit, which defines the GBZ. This is significantly different from the eigenstates with the periodic boundary condition in $x_1$, where the wavefunction is a plane wave and we have the traditional Brillouin zone with $|\beta|=1$.

In this work, we ask whether the presence of the non-Hermitian skin effect results in dynamics sensitive to the boundary condition in the thermodynamical limit. We consider the evolution of single-particle wave packets $|\psi(t)\rangle=e^{-iHt}|\psi(0)\rangle$. In the coordinate representation, we have 
\begin{equation}
\psi_{\mathbf{x},\mu}(t)=\int d\mathbf{y}~G_{\mu\nu}(\mathbf{x},\mathbf{y},t)\psi_{\mathbf{y},\nu}(0).
\end{equation}
Here $G_{\mu\nu}(\mathbf{x},\mathbf{y},t)\equiv \left<\mathbf{x},\mu\right|e^{-iHt}\left|\mathbf{y},\nu\right>$ is the single-particle Green's function. Consequently, to compare the non-Hermitian dynamics under different boundary conditions, we only need to focus on their Green's functions.

 {\color{blue}\emph{General Proof.}} -- 
The main result of this work is to establish a theorem that states $G_{\mu\nu}(\mathbf{x},\mathbf{y},t)$ is independent of the boundary condition being open or close in the thermodynamical limit, regardless of the Hamiltonian $H$ having completely different eigenenergies and eigenstates. The proof contains two steps. We first analyze the error of $G_{\mu\nu}(\mathbf{x},\mathbf{y},t)$ when truncating the infinite series of $H$ at the order of $\Lambda$, and show that it is possible to choose $\Lambda\ll L$ for obtaining an accurate estimation, where $L$ is the system size. We then focus on initial and final positions away from the boundary (with a distance larger than $\Lambda$) and prove that there is no difference between the periodic boundary condition case and the open boundary condition case. 
\vspace{10pt}

\textit{Step 1.} We first expand $e^{-iHt}$ with a cutoff $\Lambda$:
\begin{equation}\label{app}
G_{\mu\nu}^{(\Lambda)}(\mathbf{x},\mathbf{y},t)=\sum_{n=0}^\Lambda\frac{(-it)^n}{n!}\left<\mathbf{x},\mu\right|(H)^n\left|\mathbf{y},\nu\right>.
\end{equation}
We hope to estimate the difference $\delta G_{\mu\nu}(\mathbf{x},\mathbf{y},t)=G_{\mu\nu}(\mathbf{x},\mathbf{y},t)-G_{\mu\nu}^{(\Lambda)}(\mathbf{x},\mathbf{y},t)$. Inserting complete basis in the coordinate space, we have
\begin{equation} 
\delta G_{\mu\nu}= \sum_{n=\Lambda+1}^\infty\sum_{\{\alpha_i\},\{\mathbf{z}_i\}}\frac{(-it)^n}{n!}H_{\mathbf{x}\mathbf{z}_1}^{\mu\alpha_1}H_{\mathbf{z}_1\mathbf{z}_2}^{\alpha_1\alpha_2}...H_{\mathbf{z}_{n-1}\mathbf{y}}^{\alpha_{n-1}\nu}.
\end{equation}
For each summation of $\mathbf{z}_i$, there is only $(2N)^D$ non-vanishing contributions. Denoting the maximum norm of the hopping strength $t_{\mathbf{r},\mu\nu}$ as $t_{\text{max}}$, we have
\begin{equation}
|\delta G_{\mu\nu}|\leq \sum_{n=\Lambda+1}^\infty \frac{\left((2N)^Dqtt_{\text{max}}\right)^n}{n!}\approx  \frac{e^\Lambda \left((2N)^Dqtt_{\text{max}}\right)^{1+\Lambda}}{\sqrt{2\pi}\Lambda^{\frac{3}{2}+\Lambda}}.
\end{equation}
Here we have expanded for $\Lambda \gg (2N)^Dqtt_{\text{max}}$. We find the estimation $G_{\mu\nu}^{(\Lambda)}$ is accurate super-exponentially in $\Lambda$, for arbitrary system size $L$. 

\textit{Step 2.} In the thermodynamical limit, we have $L\gg \Lambda$. Considering $\mathbf{x}$ and $\mathbf{y}$ in the bulk with a distance $d\geq D\Lambda $ away from the boundary, the accurate approximation \eqref{app} contains no boundary terms. Consequently, $G^{(\Lambda)}$ is independent of the boundary condition. Finally, taking $L\rightarrow \infty$ and then $\Lambda \rightarrow \infty$, we arrive at the conclusion that the Green's function $G_{\mu\nu}(\mathbf{x},\mathbf{y},t)$ is independent of the boundary condition. 
\vspace{10pt}

Before turning to the more explicit example, we make a few comments. Firstly, we need to emphasize that what we consider is the dynamics of a bulk state at a finite time. At an infinite time or sufficiently long time for finite systems, the dynamics will eventually be influenced by boundaries, which yields a boundary-dependent relaxation time \cite{NSHEopen1}. Secondly, although here we focus on systems with translational symmetry \eqref{Hamiltonian}, the proof can be generalized to allow disorders straightforwardly when the disorder strength is bounded. 

Thirdly, we ask why the difference in eigenenergies with different boundary conditions does not lead to different Green's functions. This can be understood from the perspective of analyticity. For simplicity, we focus on $D=1$ systems with translational symmetry. Beginning from a concrete non-Hermitian system, we first impose the periodic boundary condition and then perform Fourier transformation. In the thermodynamical limit, the Green's function can thus be written as an integral over the BZ:
\begin{equation}
\begin{aligned}
G_{\mu\nu}(x,y,t)&\equiv \int \frac{dk}{2\pi}~e^{ik(x-y)}\left<\mu\right|e^{-iH(e^{ik})t}\left|\nu\right>\\
&=\oint_{|\beta|=1} \frac{d\beta}{2\pi i\beta}~\beta^{x-y}\left<\mu\right|e^{-iH(\beta)t}\left|\nu\right>.
\end{aligned}
\end{equation}
Here we have introduced $\beta=e^{ik}$. $H(\beta)_{\mu\nu}=\sum_{x=-N}^N t_{x,\mu\nu}\beta^{x}$ is the Bloch Hamiltonian and is analytic on the whole complex plane except for $\beta=0,\infty$. If the open boundary condition is taken, we expect the same expression still persists, up to a change of intergral contour to the GBZ
\begin{equation}\label{G_GBZ}
\begin{aligned}
G_{\mu\nu}(x,y,t)=\oint_{\text{GBZ}} \frac{d\beta}{2\pi i\beta}~\beta^{x-y}\left<\mu\right|e^{-iH(\beta)t}\left|\nu\right>.
\end{aligned}
\end{equation}
This expression is confirmed directly later in the non-Hermitian SSH case. One can show that $e^{-iH(\beta)t}$ is bounded by a similar estimation as the step 1, indicating it is also analytic with $\beta$, away from $\beta=0,\infty$. Consequently, the integrals over BZ and GBZ must contribute to the same result. This proves the validity of  \eqref{G_GBZ} and gives us an understanding of the boundary condition independence.

Finally, we discuss the consistency between our theorem and previous works. In  \cite{xue2020non}, authors consider the frequency space Green's function $G(x,y,\omega)$ for non-Hermitian Hamiltonians in 1D with single site per unit cell. It is found that for certain hopping parameters, $G(x,y,\omega)$ is sensitive to the boundary condition. Mathematically, the reason is that when computing $G(x,y,\omega)$, we are dealing with $(\omega-H)^{-1}$. However, the Tayor series now probably diverge for certain $\omega$, which is different from the $e^{-iHt}$ case with an infinite convergent radius. Physically, when we measure the frequency response, we are considering the limit where we take $t\rightarrow \infty$ before taking the thermodynamical limit $L\rightarrow \infty$. As a result, signals from the boundary can propagate into the bulk and lead to non-trivial effects.
	\begin{figure}[tb]
		\centering
		\includegraphics[width=1\linewidth]{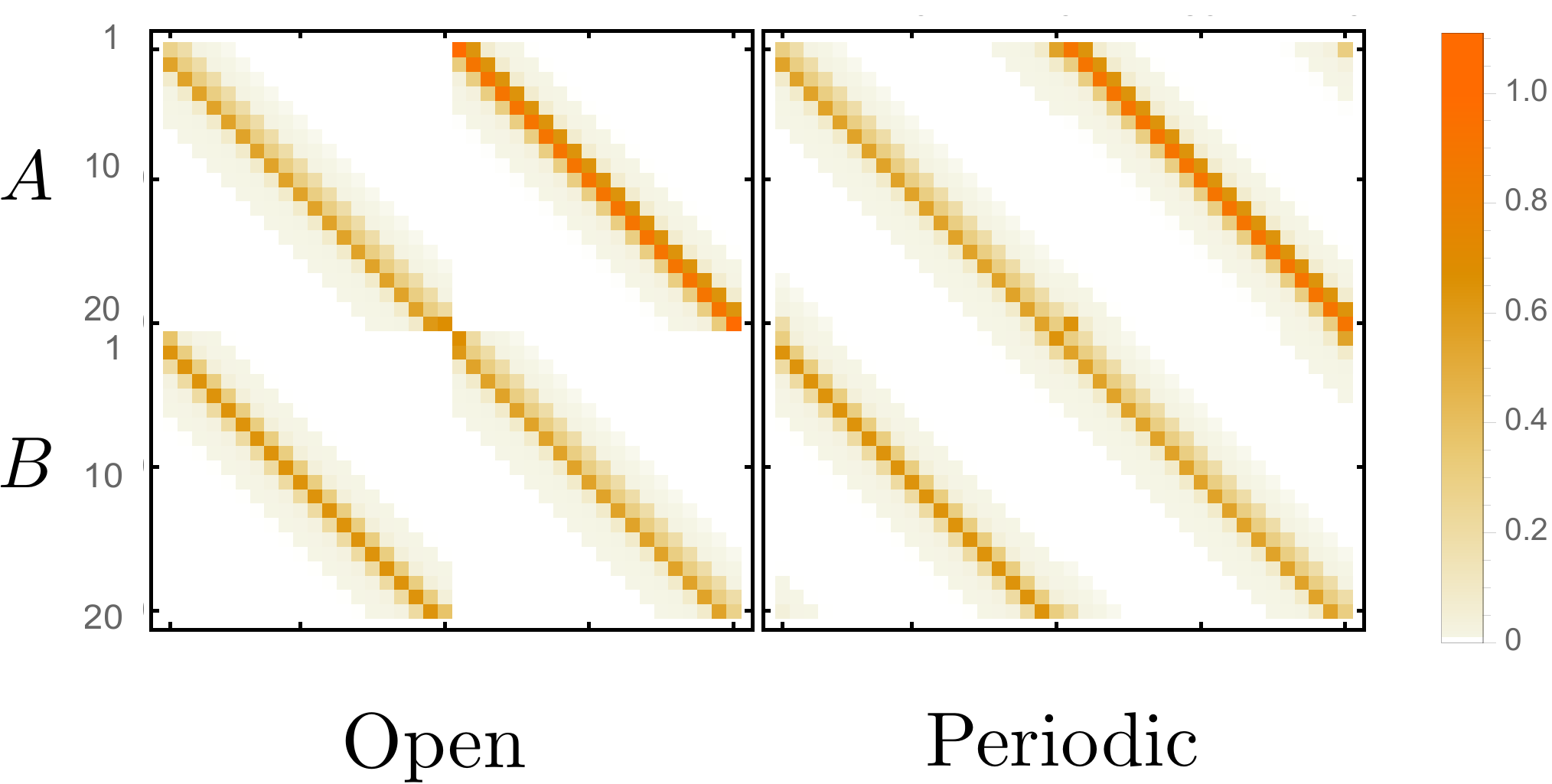}
		\caption{The numerical result for the norm of the Green's function $|G_{\mu\nu}(x,y,t)|$ for the non-Hermitian SSH model. Here we take $\gamma=t_1=t_2$ and $t\gamma=1$. From the figure, we see the Green's function in the bulk is independent of the boundary condition, as predicted by our theorem.}
		\label{fig:2}
	\end{figure}

 {\color{blue}\emph{Example: Non-Hermitian SSH Model.}} -- 
 Now we present results for the non-Hermitian SSH, where the Green's function can be derived explicitly. The Hamiltonian reads
\begin{equation}
\begin{aligned}
H=&\left(t_1+\frac{\gamma}{2}\right)\sum_x|x,A\rangle\langle x,B|+\left(t_1-\frac{\gamma}{2}\right)\sum_x|x,B\rangle\langle x,A|
\\&+t_2\sum_x|x,B\rangle\langle x+1,A|+t_2\sum_x|x+1,A\rangle\langle x,B|.
\end{aligned}
\end{equation}
This model has been widely studied in previous works \cite{NSHE1,NSHE3,nonssh1,nonssh2,nonblochband1,nonssh3}. To avoid complications we assume $t_1>\gamma/2>0$ and $t_2>0$. The Green's function can be expanded as 
\begin{equation}
G_{\mu\nu}(x,y,t)=\sum_E\langle x,\mu|E\rangle_R ~_L\langle E|y,\nu\rangle e^{-iEt}.
\end{equation} 
Here we have defined the right eigenvector $|E\rangle_{R}$ which satisfies $H |E\rangle_{R}=E|E\rangle_{R}$ and the left eigenvector $|E\rangle_{L}$ with $H^T |E\rangle_{L}=E|E\rangle_{L}$. The summation is over all eigenvectors. Under the open boundary condition, $|E\rangle_R$ takes the form
\begin{equation}
\begin{aligned}
&|E\rangle_R=\sum_n\left(\psi^A_x|x,A\rangle+\psi^B_x|x,B\rangle\right),\\
&(\psi^A_x,\psi^B_x)=\beta^x_1(\phi_1^A,\phi_1^B)+\beta^x_2(\phi_2^A,\phi_2^B),
\end{aligned}
\end{equation}
where the eigenequations in the bulk are
\begin{equation}\label{rel}
\begin{aligned}
\left(t_1+\frac{\gamma}{2}+t_2/\beta_a\right)\phi^B_a&=E\phi^A_a,\ \ \ \left(t_1-\frac{\gamma}{2}+t_2\beta_a\right)\phi^A_a=E\phi^B_a.
\end{aligned}
\end{equation}
Here $a\in\{1,2\}$. This leads to the relation 
\begin{equation}
E=\pm\sqrt{\left(t_1+\frac{\gamma}{2}+t_2/\beta_a\right)\left(t_1-\frac{\gamma}{2}+t_2\beta_a\right)}\equiv\pm E(\beta),
\end{equation}
which gives $\beta_1\beta_2=\frac{t_1-\gamma/2}{t_1+\gamma/2}$. Since bulk states require $|\beta_1|=|\beta_2|$, we introduce $\beta_1=\sqrt{\frac{t_1-\gamma/2}{t_1+\gamma/2}}e^{i\theta}=\beta_2^*$. The boundary condition is $\psi_{0}^B=0$ and $\psi_{L+1}^A=0$, or
\begin{equation}\label{bdy}
\begin{aligned}
&\phi_1^B+\phi_2^B=0,\ \ \ \ \ \beta_1^{L+1}\phi_1^A+\beta_2^{L+1}\phi_2^A=0.
\end{aligned}
\end{equation}
This leads to states labeled by $m$ as
\begin{equation}
2\theta(L+1)+2\arg\left(t_1-\gamma/2+t_2\sqrt{\frac{t_1-\gamma/2}{t_1+\gamma/2}}e^{-i\theta}\right)=2\pi m.
\end{equation}
In the thermodynamical limit $L\rightarrow \infty$, this suggests $\sum_m\rightarrow L d\theta/\pi=Ld\beta/(i\pi\beta)$. The left eigenvector $|E\rangle_L$ can be obtained by taking $\beta_a \rightarrow 1/\beta_a$ and $\gamma \rightarrow -\gamma$. Finally, we find
\begin{equation}
\begin{aligned}
G=\sum_\pm\oint_{\text{GBZ}} \frac{d\beta}{4\pi i\beta}(\beta)^{a-b}e^{\mp i E(\beta) t}\begin{pmatrix}
1& \frac{\pm E(\beta)}{t_1-\frac{\gamma}{2}+t_2\beta}\\
\frac{\pm E(\beta)}{t_1+\frac{\gamma}{2}+\frac{t_2}{\beta}} &1
\end{pmatrix}.
\end{aligned}
\end{equation}
Here we have neglected terms proportional to $\beta^L$, which averaged to zero in the thermodynamical limit. It is straightforward to check that this matches our expectation \eqref{G_GBZ}. We can also explicitly see that after summing up $\pm$, naive poles from $t_1-\frac{\gamma}{2}+t_2\beta$ or $t_1+\frac{\gamma}{2}+\frac{t_2}{\beta}$ disappear and we can deform the contour to $|\beta|=1$. Also, by deforming the integral contour, the result matches the Green's function under the periodic boundary condition.

We further provide a numerical verification for the equivalence of bulk Green's functions in FIG. \ref{fig:2}. In numerics, we take $\gamma=t_1=t_2$ and $L=20$, where there is a naive pole lying between GBZ and the traditional BZ. We directly plot the norm of the Green's function $|G_{\mu\nu}(x,y,t)|$ with open or periodic boundary conditions for $t\gamma=1$. We find that even in this case with moderate system size, the bulk Green's function is independent of the boundary condition, consistent with our theorem.

 {\color{blue}\emph{Application.}} -- 
 Our theorem can be directly applied to any system submitted to a non-Hermitian Hamiltonian. Here we give another example. By using using our theorem, we prove that the dynamics of a quantum system coupled to a Markovian bath is also independent of the boundary condition. For simplicity, we consider the 1D case with a single site per unit cell. We first prepare the system in some initial states. Then at $t=0$, the coupling to the bath is turned on. The subsequent dynamics is described by Lindblad master equation \cite{open1}
\begin{equation}
    \frac{\mathrm{d}}{\mathrm{d}t}\hat{\rho}=-i[\hat{H}_0,\hat{\rho}]+\frac{\gamma}{2}\sum_x (2\hat{L}_x\hat{\rho}\hat{L}_x^\dagger-\{\hat{L}_x^\dagger \hat{L}_x,\hat{\rho}\})
\end{equation}
If the initial state is a Gaussian state, the wick theorem works at any time $t\geq0$. Consequently, the evolution of the density matrix $\hat{\rho}$ is entirely captured by the correlation function $C_{xy}(t)=\mathrm{tr}(\hat{\rho}(t)\hat{c}_x^\dagger c_y)$. 

	\begin{figure}[tb]
		\centering
		\includegraphics[width=0.95\linewidth]{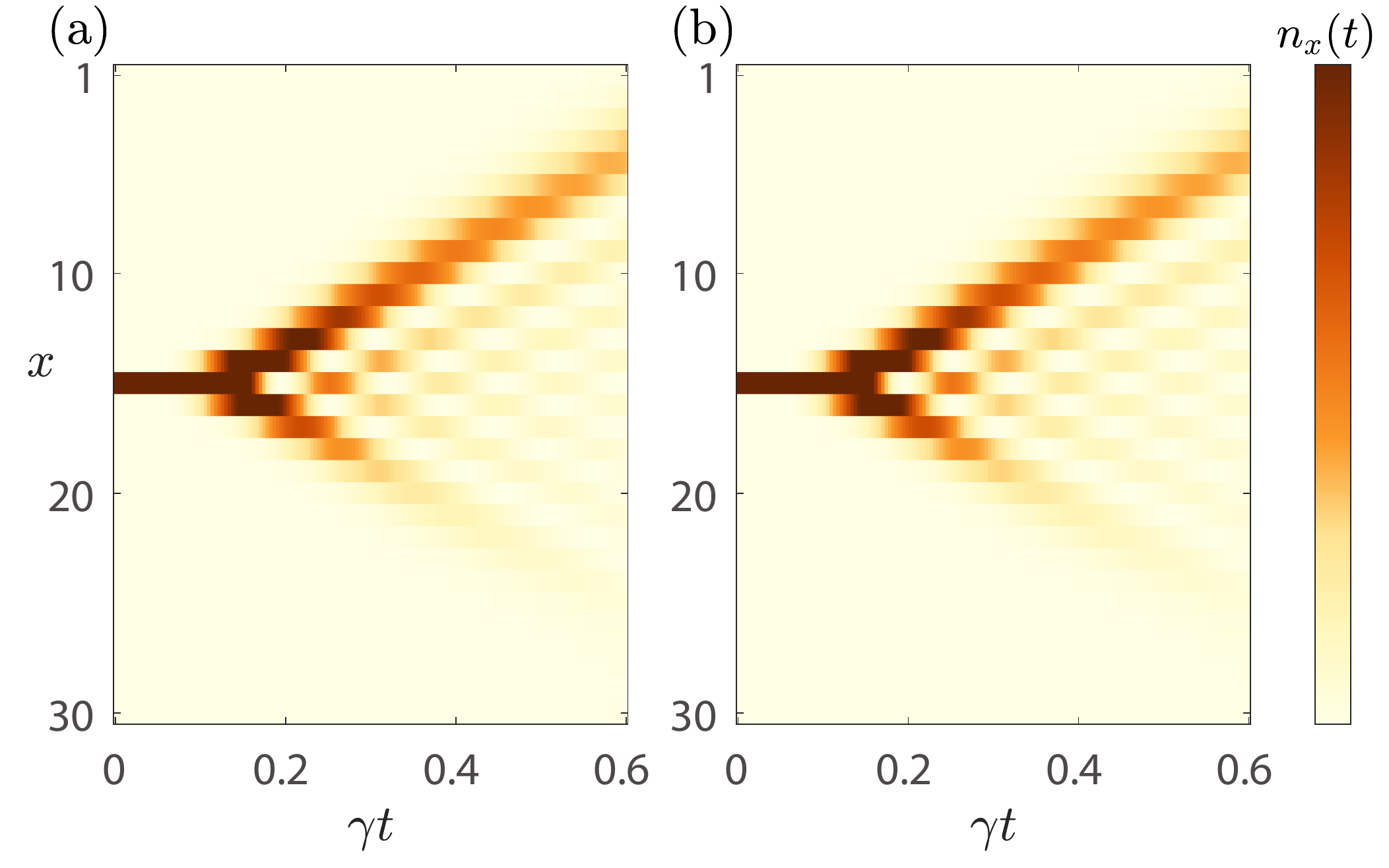}
		\caption{The dynamics of correlation function $n_x(t)=C_{xx}(t)$ in the open quantum system. The numerical results with the periodic boundary condition and the open boundary condition are shown in (a) and (b), respectively.  Here, we take $J=2u_1$,$u2=iu_1$, $\gamma=0.2 u_1$, and the system size is given by $L=30$.}
		\label{fig:3}
	\end{figure}
We further assume that the Lindblad operator $\hat{L}_x$ is a combination of annihilation operators, which take the form of $\hat{L}_x=\sum_z u_{xz}\hat{c}_z$. The dynamics of $C_{xy}(t)$ can then be represented as \cite{NSHEopen1} 
\begin{equation}
    \frac{\mathrm{d}}{\mathrm{d}t}C=-iC(H_0^T-i\frac{\gamma}{2}U)+i(H_0^T+i\frac{\gamma}{2}U)C
\end{equation}
where $C$ and $H_0$ are matrix forms of $C_{xy}$ and $(H_0)_{xy}$. $U$ is a matrix such that $U_{xy}=\sum_z u_{zx}u^*_{zy}$. As a result, the dynamics of $C$ is dominanted by propagator $G(t)=e^{i(H^T-i\frac{\gamma}{2}U)t}$,
\begin{equation}
    C(t)=G(t)C(0)G^\dagger (t)
\end{equation}
The effective Hamiltonian $H=H_0^T-i\frac{\gamma}{2}U$ is a non-Hermitian single body Hamiltonian. $G(t)$ is nothing but the single-particle Green's function. Utilizing the theorem discussed above, we can conclude that the dynamics of $C(t)$, as well as $\hat{\rho}(t)$, does not depend on boundary conditions in the thermodynamic limit. This is consistent with the fact that if we consider the system and bath as a whole, the total system is unitary and we do not expect any sensitivity to boundary conditions.

We also check our conclusion numerically with a specific model. We take the Hamiltonian $\hat{H}_0=\sum_xJ(\hat c^\dagger_{x}\hat c_{x+1}+\text{H.C.})$ and the Lindblad operators $\hat L_x=u_1 \hat c_x+u_2 \hat c_{x+1}$. At $t=0$, we prepare the system in the initial state $|\psi(0)\rangle=|x=L/2\rangle$ and $L$ is the system size.  We then compute the evolution of the correlation function $C_{xy}$ and the numerical results are shown in Fig.~\ref{fig:3}, which supports our conclusion.

 {\color{blue}\emph{Summary.}} -- 
 In this work, we generally prove that regardless of the non-Hermitian skin effect, the evolution under non-Hermitian Hamiltonians is independent of the boundary condition in the thermodynamical limit. We find that this fact can be understood by the analyticity of the integrand under the eigenstate decomposition, and an explicit example is given using the non-Hermitian SSH model. We finally discuss an application of our theorem in the evolution of open quantum systems, where we show that the evolution of the density matrix in bulk is independent of the boundary conditions. 

Having the theorem we discussed in hand, it is interesting to ask whether the non-Hermitian dynamics is independent of the boundary conditions for interacting systems. We expect that a similar theorem still exists, although proof is left for future work.

\textit{Acknowledgment.} We especially thank Hui Zhai for bring our attention to this problem and many valuable discussions during the work. We acknowledge helpful discussions with Lei Pan and Zhong Wang. PZ acknowledges support from the Walter Burke Institute for Theoretical Physics at Caltech.

\bibliography{main.bbl}
	
\end{document}